\documentclass[fleqn,usenatbib]{mnras}

\usepackage{newtxtext,newtxmath}

\usepackage[T1]{fontenc}
\usepackage{xcolor}
\definecolor{light-gray}{gray}{0.5}

\DeclareRobustCommand{\VAN}[3]{#2}
\let\VANthebibliography\thebibliography
\def\thebibliography{\DeclareRobustCommand{\VAN}[3]{##3}\VANthebibliography}

\usepackage{graphicx}	
\usepackage{amsmath}	
\usepackage{array}	%

\title[Temperature properties in magnetised accretion flows]{Temperature
  properties in magnetised and radiatively cooled two-temperature
  accretion flows onto a black hole}

\author[Dihingia et al.]{
Indu K. Dihingia,$^{1,2,3}$\thanks{E-mail: idihingia@iiti.ac.in}
Yosuke Mizuno,$^{2,4,5}$\thanks{E-mail: mizuno@sjtu.edu.cn}
Christian M. Fromm,$^{6,5,7}$
and Luciano Rezzolla$^{5,8,9}$
\\
$^{1}$Department of Astronomy, Astrophysics and Space Engineering, Indian Institute of Technology Indore, Khandwa Road, Simrol 453552, India\\
$^{2}$Tsung-Dao Lee Institute, Shanghai Jiao-Tong University, 520 Shengrong Road, Shanghai, 201210  People's Republic of China\\
$^{3}$Department of Physics, Indian Institute of Science, Bangalore 560012, Karnataka, India\\
$^{4}$School of Physics \& Astronomy, Shanghai Jiao-Tong University, 800 Dongchuan Road,  Shanghai, 200240, People's Republic of China\\
$^{5}$Institut f\"ur Theoretische Physik, Goethe Universit\"at, Max-von-Laue-Str. 1, 60438 Frankfurt am Main, Germany\\
$^{6}$Institut f\"ur Theoretische Physik und Astrophysik, Julius Maximilian University, W\"urzburg, Emil-Fischer-Str. 31, D-97074 W\"urzburg, Germany\\
$^{7}$Max-Planck-Institut f\"ur Radioastronomie, Auf dem H\"ugel 69, D-53121 Bonn, Germany\\
$^{8}$School of Mathematics, Trinity College, Dublin 2, Ireland\\
$^{9}$Frankfurt Institute for Advanced Studies, Ruth-Moufang-Str. 1, 60438 Frankfurt am Main, Germany
}

\date{Accepted XXX. Received YYY; in original form ZZZ}

\pubyear{2015}

\begin{document}
\label{firstpage}
\pagerange{\pageref{firstpage}--\pageref{lastpage}}
\maketitle

\begin{abstract}
Simplified assumptions about the thermodynamics of the electrons are
normally employed in general-relativistic magnetohydrodynamic (GRMHD)
simulations of accretion onto black holes. To counter this, we have
developed a self-consistent approach to study magnetised and radiatively
cooled two-temperature accretion flows around a Kerr black hole in two
spatial dimensions. The approach includes several heating processes,
radiative cooling, and a coupling between the electrons and the ions via
Coulomb interaction. We test our approach by performing axisymmetric
GRMHD simulations of magnetised tori accreting onto a Kerr black hole
under various astrophysical scenarios. In this way, we find that the
inclusion of the Coulomb interaction and the radiative cooling impacts
the thermodynamical properties of both the ions and electrons, changing
significantly the temperature distribution of the latter, and underlining
the importance of a two-temperature approach when imaging these flows. In
addition, we find that the accretion rate influences the bulk properties
of the flow as well as the thermodynamics of the electrons and
ions. Interestingly, we observe qualitatively distinct temperature
properties for SANE and MAD accretion modes while maintaining the same
accretion rates, which could help distinguishing MAD and SANE accretion
flows via observations. Finally, we propose two new relations for the
temperature ratios of the electrons, ions, and of the gas in terms of the
plasma-$\beta$ parameter. The new relations represent a simple and
effective approach to treat two-temperature accretion flows on
supermassive black holes such as Sgr A* and M\,87*.
\end{abstract}




\section{Introduction}

Already in 1976, \cite{Shapiro-etal1976} (hereafter SLE) proposed
two-temperature accretion flows as a way to explain the observed spectra
from Cyg X-1, where electrons and ions are regarded as two separate
fluids with different temperatures. The SLE model has the benefit
of a ``hotter" solution for the accretion flow than the conventional
\cite{Shakura-Sunyaev1973} thin-disc model, which is plausible 
to explain observed hard X-rays in Cyg X-1. Soon after,
the SLE model was used to investigate the spectral characteristics of
X-Ray Binaries (XRBs) and Active Galactic Nuclei (AGNs) (e.g.,
\cite{Ichimaru1977, Rees-etal1982, Kusunose-Takahara1988,
  Kusunose-Takahara1989, White-Lightman1989, Wandel-Liang1991,
  Luo-Liang1994}. In addition, the SLE model was modified to include
advection terms which remove the inherently thermally unstable nature of
the model \citep{Pringle1976,Piran1978}, and it is explored in a wide
range of physical settings using a semi-analytic technique
\citep[etc.]{Abramowicz-etal1988, Kato-etal1988, Narayan-Yi1995,
  Esin-etal1996, Nakamura-etal1996, Manmoto-etal1997,
  Rajesh-Mukhopadhyay2010, Dihingia-etal2018, Dihingia-etal2020,
  Sarkar-etal2020}. While these semi-analytical models are crucial
because they give a foundation for understanding the accretion processes
around black holes in greater detail. However, numerical simulations are
necessary to depict the realistic time-dependent, turbulent nature of the
accretion flow.

General-relativistic magnetohydrodynamic (GRMHD) simulations are
typically performed in a single-fluid or single-temperature approximation
to comprehend emission near the black hole \citep[e.g.,
][]{Noble-etal2007, Moscibrodzka-etal2009, Moscibrodzka-etal2012,
  Moscibrodzka-etal2014, Moscibrodzka-etal2016, Dexter-etal2009,
  Dexter-etal2010, Shcherbakov-etal2012, Chan-etal2015, Gold-etal2017,
  Porth-etal2017, Mizuno-etal2018, Davelaar-etal2018,
  Davelaar-etal2019}. The single-fluid approximation does not treat the
temperatures of the electrons and ions self-consistently. As a result,
the electron temperature required to calculate the electromagnetic
emission needs to be estimated in some way from the gas temperature, which is
computed self-consistently in the GRMHD simulations. Obviously, this
estimate requires some suitable assumption from which to derive a
prescription between the temperature of the electrons $T_e$ and that of
the ion/gas $T_{i/g}$.

These prescriptions are either very simple, e.g., a constant rescaling of
the gas temperature \citep{Moscibrodzka-etal2009, Moscibrodzka-etal2014,
  Chan-etal2015} or employ parametric models making use of the
plasma-$\beta$ distribution \citep{Moscibrodzka-etal2016,
  Anantua-etal2020}. One of the most commonly employed prescriptions for
the electron temperature is the so-called $R-\beta$ prescription
\citep{Moscibrodzka-etal2016}, where the electron temperature is
estimated from plasma-$\beta$ with two parameters ($R_{\rm low}, R_{\rm
  high}, \beta_{\rm crit}$). This prescription was utilised to develop
synthetic emission maps to model the EHT observations of 
M\,87* and Sgr A* \citep{EHTCV,EHTCVIII,EHT-SGRAV-2022}. 
Often, a large number of these parameters ($R_{\rm low},
R_{\rm high})$ need to be tested to find a suitable combination/range of
parameters that match the observations \citep[e.g.,][]{Mizuno-etal2021,
  Fromm-etal2022,Cruz2022}. However, since the correlations of these
parameters with the physics involved in the accretion flow are not known,
it is difficult to translate the knowledge of the optimal parameters into the
knowledge of the physical nature of the flow. Therefore, a physically
motivated self-consistent two-temperature framework is important in order
to obtain firm information from the observations, and this is one of the
primary goals of this study.

\cite{Ressler-etal2015} have performed two-temperature GRMHD simulations
to incorporate electron thermodynamics self-consistently and applied them
to understand the images and spectra of Sgr A*
\citep{Ressler-etal2017}. Their models produce sufficiently hot electrons
that match the observed luminosity in NIR (near-infrared) bend.  However,
in their formalism, they only considered the effects of electron heating
but neglected the radiative cooling and Coulomb interaction. Recently,
\cite{Dexter-etal2020a} performed an extensive parameter survey of
two-temperature GRMHD simulations of magnetised accretion flows
considering black-hole spins and different electron-heating prescriptions
for Sgr A*. The study has explained the recently observed mean Faraday
rotation and the polarised signals of NIR flares in Sgr A*. With the same
numerical approach, \cite{Yao-etal2021} reported comparable luminosity of
very high-energy flares from M\,87. However, they could not able to
explain the timing properties of very high-energy flares in M\,87.
Subsequently, \cite{Sadowski-etal2017} using an M1 approximation for
  the radiative transport, and \cite{Ryan-etal2017} using a
  Montecarlo scheme extended the formalism to incorporate effects from
radiative feedback.  Improved versions of this formalism has been
recently applied to examine the images and timing properties of Sgr A* as
well as M\,87*
\citep{Chael-etal2018,Ryan-etal2018,Chael-etal2019}. Similar formalism is
also applied to black hole X-ray binaries by \cite{Dexter-etal2021}, they
studied accretion flows with different accretion rates around a
stellar-mass black hole and found the collapse of hot accretion flows due
to the thermal runaway.

Also, it is worth mentioning that single-fluid GRMHD with radiative
cooling also has been explored to study hot accretion flow for Sgr A*
\citep[e.g.,][]{Fragile-Meier2009, Dibi-etal2012, Drappeau-etal2013,
  Yoon-etal2020}. According to \cite{Yoon-etal2020}, the role of
  radiative cooling processes in the dynamics of the accretion flows
  increases with the accretion rate. They also identified a critical
  accretion rate that is $\dot{M}\gtrsim10^{-7}$ times the Eddington
  limit; beyond this value, radiative cooling becomes dynamically
  important However, in these works, the radiation-cooling rate is
calculated from an adjusted electron temperature that assumes the ratio
of electron to ion/gas temperature to be constant ($T_e/T_i=$
constant). Such consideration is not realistic and may introduce
artificial effects in the underlying conclusions from these studies.

The goal of this study is to offer a formalism to obtain information on
electron thermodynamics using a self-consistent two-temperature
paradigm. To test this formalism, we perform axisymmetric 2D GRMHD
simulations around a Kerr black hole. In this way, we investigate the
transfer of energy from the ions to the electrons via the
Coulomb-interaction process, which serves as a heating mechanism for the
electrons and as a cooling mechanism for the ions. Additionally,
electrons also gain energy through the turbulent heating process
\citep{Howes2011}. Finally, electrons can lose energy by several
radiative mechanisms, such as Bremsstrahlung, synchrotron radiation, and
the inverse Compton process. We illustrate the temperature distributions
($T_e, T_i$) in several physical scenarios and show the correlation of
different temperatures and plasma-$\beta$ from the simulation
models. Future research will examine how these findings impact emission,
spectra, and images computed with general-relativistic ray-tracing (GRRT)
calculations.

The rest of the paper is organised as follows: Section \ref{sec:two}
focuses on numerical and mathematical formalism, while
Sec. \ref{sec:three} discusses the outcomes of our simulation
models. Finally, Sec. \ref{sec:four} summarises and discusses our
findings, as well as our future plans. Note that the Roman indices run
from $1$ to $3$, while the Greek indices run from $0$ to $3$. Further,
all the equations are solved in the code units, where $G$ (gravitational
constant) $=M_{\bullet}$ (mass of the central black hole) $=c$
(speed of light) $= 1$. Subsequently, we express the unit of mass,
length, and time in terms of $M_{\bullet}$, $GM_{\bullet}/c^2$, and
$GM_{\bullet}/c^3$, respectively.

%

\section{Mathematical and physical setup}
\label{sec:two}

For our simulations, we have employed the adaptive-mesh refinement (AMR)
GRMHD code \texttt{BHAC} \citep{Porth-etal2017,Olivares-etal2019} to
perform 2D simulations of magnetised tori around a Kerr black hole. The
ideal GRMHD equations can be expressed in terms of the conservation laws
as follows \citep[see][]{DelZanna-etal2007, Rezzolla-Zanotti2013},
\begin{align}
    \partial_t\left(\sqrt{\gamma}\boldsymbol{U}\right) + \partial_i\left(\sqrt{\gamma}\boldsymbol{F^i}\right) = \sqrt{\gamma}\boldsymbol{S}\,.
    \label{eq:01}
\end{align}
In the presence of radiative cooling, we modify the source term as
$\boldsymbol{ S}=\boldsymbol{S_0} + \boldsymbol{S^\prime}$, where
$\boldsymbol{S^\prime}$ is the additional contribution from
radiation-cooling processes. The explicit expressions for the vectors
of conserved variables $\boldsymbol{U}$, fluxes $\boldsymbol{F^i},$
  and sources $\boldsymbol{S_0}$ in the case of ideal GRMHD without
  radiative cooling have been reported by \cite{Porth-etal2017} [see
    Eqs. (23) and (30) there]. The vector of conserved variables is given
  by $\boldsymbol{U}\equiv[{\cal D}, {\cal S}_j, \tau, {\cal B}_j]^{T}$,
  where ${\cal D}, {\cal S}_j, \tau,$ and $ {\cal B}_j$ refer to density,
  covariant three-momentum, the rescaled energy density, and magnetic
  three-fields in Eulerian frame, respectively \citep[for detail
    follow][]{Rezzolla-Zanotti2013}. On the other hand, the explicit
form of the new source term $\boldsymbol{S^\prime}$ can be expressed as
follows
\begin{align}
    \boldsymbol{S^\prime}=
    \begin{pmatrix}
0 \\
-\alpha \gamma v_j \Lambda \\
-\alpha \gamma \Lambda \\
0 \\
\end{pmatrix}\,,
\label{eq:02}
\end{align}
where $\alpha, \gamma, v_j$, and $\Lambda$ correspond to the
lapse-function, Lorentz factor, fluid three-velocity, and the total
radiation-cooling term, respectively. Note that here we consider the
  radiation-flux term to be proportional to the radiative cooling rate
  and the fluid velocity following \cite{Fragile-Meier2009,Dibi-etal2012,
    Yoon-etal2020}; both choices represent simplifications over
  a more accurate M1 moment approach closure
  \citep[e.g.,][]{Sadowski-etal2017} or Montecarlo scheme
  \citep[e.g.,][]{Ryan-etal2017}.

Equation \eqref{eq:01} solves the single-fluid MHD equations containing
electrons and ions. To extend our study to a two-temperature
framework, we additionally solve the electron-entropy equation in the
presence of dissipative heating, Coulomb interaction, and
radiation-cooling processes, which is given by \citep[see,
  e.g.,][]{Sadowski-etal2017}
\begin{align}
    T_e \nabla_\mu \left(\rho u^\mu \kappa_e \right) = f Q + \Lambda_{\rm ei} - \Lambda\,,
\label{eq:03}
\end{align}
where $\kappa_e := \exp[(\tilde{\Gamma}_e -1)s_e]$ and
  $s_e:=p/\rho^{\tilde{\Gamma}_e}$ is the electron entropy per particle.
Here, $\tilde{\Gamma}_e$ is the adiabatic index of the electrons, $Q$ is
the rate of dissipative heating, and $f$ is the fraction of dissipative
heating transferred to the electrons. $\Lambda_{\rm ei}$ is the energy
transferred to the electrons from the ions due to the Coulomb
interaction. We also assume charge neutrality, i.e., $n=n_e=n_i$
  (equal number densities for electrons and ions), and the
four-velocities of the flow to be the same as the four-velocities of
electrons and ions, i.e., $u^\mu=u^\mu_e=u^\mu_i$ when solving the
electron-entropy equation. Here, we treat the dissipative-heating term
following \cite{Ressler-etal2015} and subsequently update the electron
entropy explicitly following the conservation laws for electrons.

\subsection{Dissipative and radiative processes}

By solving Eqs. \eqref{eq:01} and \eqref{eq:03}, we obtain all the
properties of the flow and of the electrons. Using these properties, we
estimate the internal energy density of the ions from the total
internal energy density of the flow, and they are related as
\begin{align}
    u_{g, {\rm int}} = u_{i, {\rm int}} + u_{e, {\rm int}}\,, 
\end{align}
where the subscripts $g, i$, and $e$ stand for the gas (meant here as the
global flow), ions, and electrons, respectively. We assume an
ideal-fluid equation of state \citep[see, e.g.,][]{Rezzolla-Zanotti2013}
to calculate the internal energies, i.e., $u_k=p_k/(\tilde{\Gamma}_k -
1)$, where $p_k$ and $\tilde{\Gamma}_k$ are the isotopic pressure and the
adiabatic index of the species ($k=g,i,e$). In a realistic scenario, the
temperatures of the components are different in different parts of the
simulation domain and also vary during the dynamic evolution of the
flow. Consequently, the adiabatic indices should also change following
the relativistic equation of state \citep{Chandrasekhar1939,
  Synge1957}. For simplicity, we consider constant adiabatic indices that
are constant in space and time for the electrons, ions, and the gas as
a whole. Indeed, considering temperature-dependent adiabatic indexes,
\cite{Sadowski-etal2017} have shown that $\tilde{\Gamma}_e$ is always of
the order of $\sim 4/3$ and that since the ion temperature is normally
$T_i\lesssim10^{11.5}$K, -- so that the ions can be considered as
non-relativistic -- it is reasonable to assume
$\tilde{\Gamma}_i\sim5/3$. Also, since \cite{Sadowski-etal2017} report
that the adiabatic index of the flow varies between $5/3$ (in the disk
midplane) and $4/3$ (in the polar region), we decide to set
$\tilde{\Gamma}_e=4/3$ and $\tilde{\Gamma}_i=5/3$, and
$\tilde{\Gamma}_g=13/9$ as for a gas that is thermally trans-relativistic
\citep[e.g.,][]{Shiokawa-etal2012,Ryan-etal2017,Ryan-etal2018}.

From the pressure of the individual components, we can calculate the
corresponding temperatures $T_k$ assuming an ideal-gas law, i.e.,
$p_k=\rho k_{\rm B}T_k/m_k$, where $k_{\rm B}$ is the Boltzmann constant
and $m_k$ is the mass of $k$-th components of the flow. Using these
temperatures and other flow variables, we can calculate the
dissipative-heating fraction, the Coulomb-interaction rate, and
radiation-cooling rates as detailed below. More specifically, we consider
a turbulent-heating prescription to determine the fraction of dissipative
heating $f$. To model this quantity, we follow the results of numerical
simulations by \cite{Howes2010,Howes2011}. In particular, the fraction $f$ is
simply defined as
\begin{align}
    f := \frac{1}{1+Q_i/Q_e}\,,
    \label{eq:05}
\end{align}
where
\begin{align*}
    \frac{Q_i}{Q_e} = c_1 \frac{c_2 + \beta^\alpha}{c_2 + \beta^\alpha}
    \sqrt{\frac{m_p T_i}{m_eT_e}}\exp(-1/\beta)\,,
\end{align*}
where $m_e$ and $m_p$ refer to the mass of electron and proton,
respectively, while $c_1=0.92, c_2=1.6/(T_i/T_e), c_3 = 18 + 5
\log(T_i/T_e)$, $\alpha=2-0.2\log(T_i/T_e)$, and $\beta=p_{\rm
  gas}/p_{\rm mag}$. Next we express the Coulomb interaction and
  radiation-cooling terms used in Eqs.~\eqref{eq:02}--\eqref{eq:03} in
  CGS units. Finally, we convert these terms to the units adopted by the
  code before returning to Eq.~\eqref{eq:02}--\eqref{eq:03} as
  $\Lambda=Q^-/U_{\rm c}$ and $\Lambda_{\rm ei}=Q_{\rm ei}/U_{\rm c}$,
  where $U_{\rm c}=\dot{M}_{\rm cgs}c^2/r_g^3$ and $\dot{M}_{\rm cgs}$ is
  the mass-accretion rate $(\dot{M})$ in CGS units. Accordingly, we
express the Coulomb-interaction rate $(Q_{\rm ei})$ following
\cite{Spitzer1965, Colpi-etal1984}, whose explicit form in CGS units is
given by
\begin{align}
    Q_{\rm ei} = 1.6\times10^{-13} \frac{k_{\rm
        B}\sqrt{m_e}\ln\Lambda_0}{m_p}n_e n_i (T_i - T_e) T_e^{-3/2}\,,
\label{eq:cc}
\end{align}
where we consider $\ln\Lambda_0=20$.

In order to calculate the Bremsstrahlung-cooling rate, we follow the
prescriptions reported by \cite{Esin-etal1996}. More precisely, the
free-free Bremsstrahlung-cooling rate for an ionised plasma consisting of
electrons and ions is given by $Q_{\rm br}=Q_{\rm br}^{\rm ei} +
Q_{\rm br}^{\rm ee}$, where the explicit forms of the individual terms
are
\begin{align}
\begin{aligned}
    Q_{\rm br}^{\rm ei} = & 1.48\times10^{-22} n_i n_e\\ & \times 
    \begin{cases}
    4\sqrt{\frac{2\Theta_e}{\pi^3}}\left(1+1.781\Theta_e^{1.34}\right)\,,&
    \text{if } \Theta_e <
    1\\ \frac{9\Theta_e}{2\pi}\left(\ln\left(1.123\Theta_e + 0.48\right)
    + 1.5\right)\,, & \text{otherwise.}
  \end{cases}
\end{aligned}
\end{align}
and
\begin{align}
\begin{aligned}
   & Q_{\rm br}^{\rm ee} = \\ &
    \begin{cases}
    2.56\times10^{-22}n_e^2 \Theta_e^{3/2}\left(1 + 1.1\Theta_e +
    \Theta_e^2 - 1.25\Theta_e^{5/2}\right)\,,& \text{if } \Theta_e <
    1\\ 3.24\times10^{-22} n_e^2 \Theta_e \left(\ln(1.123\Theta_e) +
    1.28\right)\,, & \text{otherwise.}
    \end{cases}
\end{aligned}
\end{align}
Here, $\Theta_e:=k_{\rm B}T_e/m_e c^2$ is the dimensionless electron
temperature.

\begin{figure}
    \centering
    \includegraphics[scale=0.6]{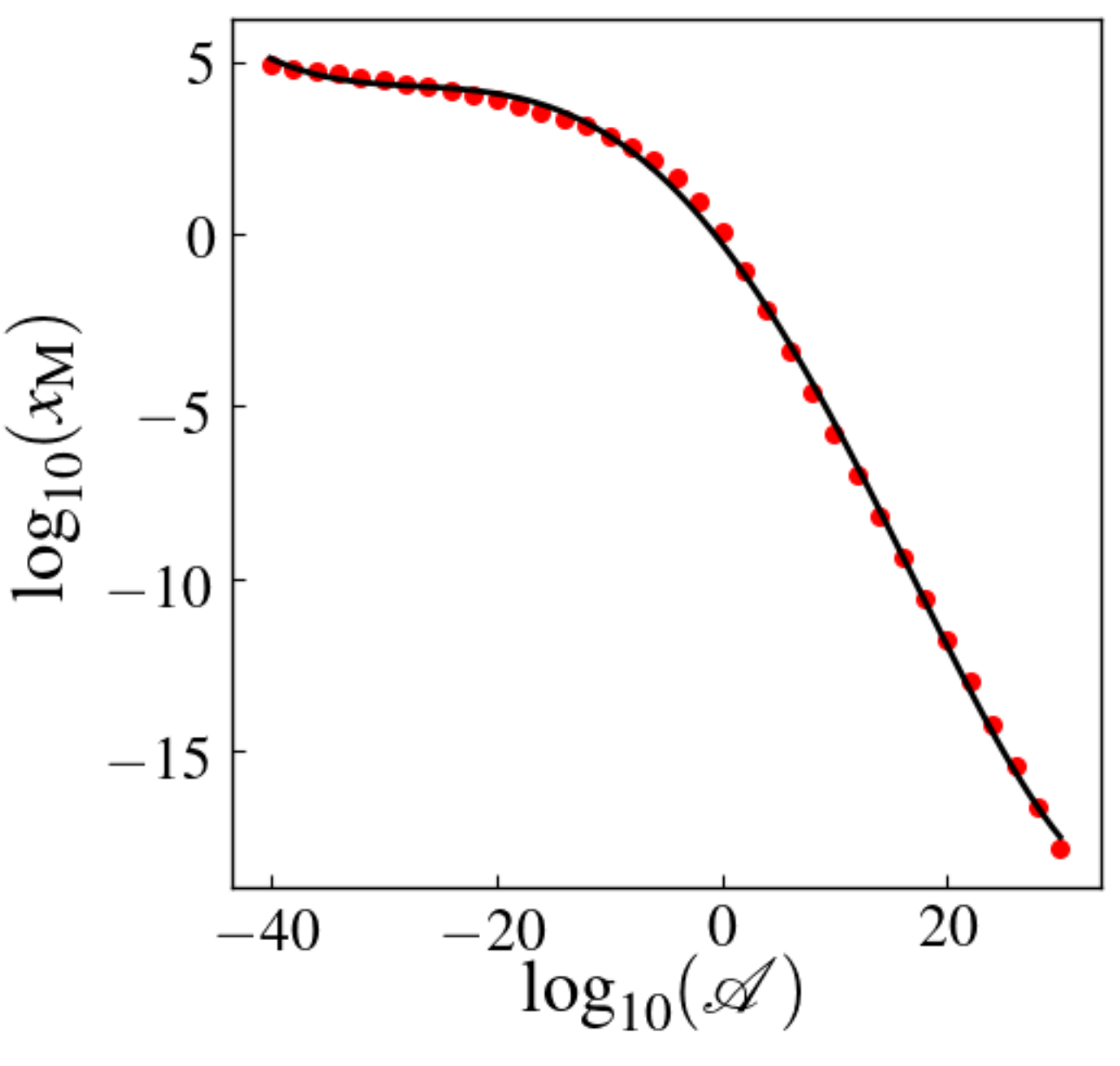}
    \caption{Plot of quantity $x_{\rm M}:=2\nu/(3\nu_0\Theta_e^2)$
      appearing in Eq.~\eqref{eq:calA} shown as a function of the variable
      ${\cal A}$. The numerical solution is shown by the red dots line
      while its polynomial fit is shown with the black solid
      line.}
    \label{fig:xm}
\end{figure}

Because of the presence of a strong magnetic field, the hot electrons in
the accretion flow radiate via the thermal synchrotron process. We
consider the rate of synchrotron emission \citep{Esin-etal1996} as
follows
\begin{align}
  \begin{aligned}
    Q_{\rm cs} = & \frac{2\pi k_{\rm B} T_i \nu_c^3}{3Hc^2} + 6.76\times10^{-28}\frac{n_e}{K_2(1/\Theta_e)a_1^{1/6}}\\
                 & \times\bigg[\frac{1}{a_4^{11/2}}\Gamma\left(\frac{11}{2},a_4\nu_c^{1/3}\right) + \frac{a_2}{a_4^{19/4}}\Gamma\left(\frac{19}{4},a_4\nu_c^{1/3}\right) \\
                 &+ \frac{a_3}{a_4^4}\left(a_4^3 \nu_c + 3a_4^2\nu_c^{2/3} + 6a_4\nu_c^{1/3} +6\right)\exp(-a_4\nu_c^{1/3})\bigg]\,,\\
\end{aligned}
\label{eq:cool2}
\end{align}
where $H$ is the local scale-height, which we estimate from the gradient
of the electron temperature is calculated by $H=T_e^4/|\nabla T_e^4|$
\citep{Fragile-Meier2009} and finally, $K_2$ is the modified
Bessel function of the second kind. The coefficients $a_{1-4}$ in expression
\eqref{eq:cool2} have explicit expressions
\begin{align}
  a_1=\frac{2}{3\nu_0\Theta_e^2}\,, \quad
  a_2=\frac{0.4}{a_1^{1/4}}\,, \quad
  a_3=\frac{0.5316}{a_1^{1/2}}\,, \quad
  a_4 = 1.8899a_1^{1/3}\,,
\end{align}
while $\Gamma$ is the ``Gamma function'', namely,
$\Gamma(a,x):=\int_x^\infty t^{a-1}e^{-t}~dt$. Additionally,
$\nu_0:=eB/(2\pi m_e c)$ and $\nu_c$ are the characteristic synchrotron
frequencies, where $e$ and $B$ refer to the electronic charge and
  the magnetic field strength in CGS units, respectively. The
  characteristic synchrotron frequencies can be calculated by equating
the emissivities of optically thin and thick volumes
\citep{Esin-etal1996},
\begin{align}
    \frac{e^2}{c\sqrt{3}}\frac{4\pi \nu_c n_e
    }{K_2(1/\Theta_e)}I^\prime(x_{\rm M})=\frac{2\pi k_{\rm B}T_e}{H
      c^2}\nu_c^2\,,
    \label{eq:syn}
\end{align}
with
\begin{equation}
I^\prime(x_{\rm M}) = \frac{4.05}{x^{1/6}_{\rm M}}
\left(1+\frac{0.4}{x^{1/4}_{\rm M}}+\frac{0.5316}{x^{1/2}_{\rm M}}\right)
\exp \left(-1.8899 x^{1/3}_{\rm M}\right)\,,
\end{equation}
and where $x_{\rm M}:=2\nu_c/(3\nu_0\Theta_e^2)$.  For simplicity, we
rewrite Eq.~\eqref{eq:syn} as
\begin{align}
    I^\prime(x_{\rm M})={\cal A}~x_{\rm M}\,,
    \label{eq:syn1}
\end{align}
where
\begin{align}
  \label{eq:calA}
  {\cal A} := \frac{2 \sqrt{3}\pi m_e c \Theta_e K_2(1/\Theta_e)}{4 e^2
      n_e H}\frac{3\nu_0\Theta_e^2}{2} \sim \frac{3\sqrt{3}}{4} \frac{m_e
      c}{e^2}\frac{\Theta_e^5\nu_0}{n_e H}\,,
\end{align}
and where we consider $K_2(1/\Theta_e) \sim \Theta_e^2.$ We numerically solve
Eq.~\eqref{eq:syn1} for $x_{\rm M}$ for all possible values of ${\cal A}$
within the range $[10^{-40},10^{30}]$. Subsequently, we fit the solutions
with a polynomial of fifth-order as follows
\begin{align}
\begin{aligned}
    \log_{10}(x_{\rm M}) & = A_0+ A_1 \log_{10}({\cal A})+ A_2
    (\log_{10}({\cal A}))^2\,,\\ &+ A_3 (\log_{10}({\cal A}))^3 + A_4
    (\log_{10}({\cal A}))^4 + A_5 (\log_{10}({\cal A}))^5\,,\\
\end{aligned}
\end{align}
where $A_0 = -0.273264, A_1 = -0.421236, A_2 = -0.0108989, A_3 = 4.24618
\times 10^{-5}, A_4 = 4.33030 \times 10^{-6},~{\rm and}~ A_5 = 2.35786
\times 10^{-8}.$ The comparison of the numerical solutions (dots) and the
polynomial fit (solid line) is shown in Fig.~\ref{fig:xm} and clearly
indicates the quality of the fit. Interestingly, this approach
effectively reduces the computational costs of about four orders of
magnitude.

It is important to remark that Eq.~\eqref{eq:cool2} is appropriate only
in the presence of thermal electrons. In order to ensure that the
contribution to the thermal synchrotron radiation from the
highly-magnetised region is negligible, we modify \eqref{eq:cool2} with
the help of a cutoff value for the magnetisation $\sigma$, namely, the
ratio between rest-mass and magnetic energy densities: $\sigma :=
b^2/\rho$. More specifically, we set $Q^\prime_{\rm
  cs}=\exp(-(\sigma/\sigma_{\rm cut})^2)Q_{\rm cs}$ with $\sigma_{\rm
  cut}=10$. 

In addition, we also incorporate the Comptonization of synchrotron
  radiation. In our simplified model, we calculate the
  Compton-enhancement factor of the synchrotron radiation at the local
  cuff off frequency $(\nu_c)$. In this way, the total radiation-cooling
rate is calculated as
\begin{align}
    Q^-=Q_{\rm br} + \eta(\nu_c) Q^\prime_{\rm cs}\,,
\end{align}
where $\eta$ is the Compton enhancement factor, whose explicit expression
is \citep{Narayan-Yi1995}
\begin{align}
    \eta(\nu_c) = 1 + \frac{P(A-1)}{1-PA}\left(1 - \left(\frac{3k_{\rm
        B}T_e}{h \nu_c}\right)^{\eta_1}\right)\,,
\end{align}
with, $P = 1-\exp(-\tau_{\rm es})$, $A = 1 + 4\Theta_e + 16\Theta_e^2$,
$\eta_1 = 1 + \ln P/\ln A$, and $\tau_{\rm es}=2\sigma_{\rm T}n_e H$,
where $\sigma_{\rm T}$ is the Thomson cross section of the
  electron.

\subsection{Model setup}

\begin{table}
\centering
\setlength{\tabcolsep}{.3em}
\begin{tabular}{| l  l  c |}
    \hline
    \hline
    Model & \phantom{Turbulent} Physical processes & $\dot{M}/\dot{M}_{\rm Edd}$\\ 
          &                                        & $\log_{10}$\\ 
    \hline
    \hline
    \texttt{T}      & Turbulent heat. & $-4$\\
    \texttt{TC}     & Turbulent heat. + Coulomb heat. & $-4$ \\
    \texttt{TCR-A}  & Turbulent heat. + Coulomb heat. + Radiative cool. & $-4$ \\
    \texttt{TCR-B}  & Turbulent heat. + Coulomb heat. + Radiative cool. & $-5$ \\
    \texttt{TCR-C}  & Turbulent heat. + Coulomb heat. + Radiative cool. & $-6$ \\
    \texttt{TCR-D}  & Turbulent heat. + Coulomb heat. + Radiative cool. & $-8$\\
    \texttt{TCR-DM} & Turbulent heat. + Coulomb heat. + Radiative cool. & $-8$\\
    \texttt{TCR-E}  & Turbulent heat. + Coulomb heat. + Radiative cool. & $-11$\\
    \hline
  \end{tabular}
\caption{List of the various models considered in the simulations
  alongside with the microphysical processes that are taken into
  account. Also reported for the different models are the accretion rates
  expressed in terms of the Eddington accretion rate $\dot{M}_{\rm
    Edd}$.}
\label{tab-01}
\end{table}

We initialise our simulations of magnetised tori with an axisymmetric
torus in pressure equilibrium \citep{Fishbone-Moncrief1976} with
parameters $r_{\rm in} = 6 r_g$ and $r_{\rm max} = 12 r_g$, where $r_{\rm
  in}$ and $r_{\rm max}$ are the inner radius and the radius of the
pressure maximum of the torus, respectively. Given this combination of
parameters, the subsequent flow leads to a SANE accretion flow. In a SANE
flow, the magnetic field strength is weak and merely serves as a
mechanism to transport angular momentum \citep[e.g.,][]{Narayan-etal2012,
  Porth-etal2017, Nathanail-etal2020}. However, for completeness, we have
also considered a MAD accretion modes (i.e., model \texttt{TCR-DM}),
which is triggered by a torus with $r_{\rm in} = 20 r_g$ and $r_{\rm max}
= 40 r_g$. In a MAD flow, by contrast, the magnetic field is strong
enough to govern the dynamics of the flow
\citep[e.g.,][]{Tchekhovskoy-etal2011, Narayan-etal2012}. In all cases,
the magnetic field is introduced via a weak poloidal
single-loop\footnote{We note that a multi-loop initial magnetic-field
configuration has also been considered and shown to lead to a different
and interesting phenomenology that has potential applications in
understanding flaring activities in the supermassive black hole, e.g.,
Sgr A* \citep{Nathanail-etal2020}} magnetic field specified via a vector
potential $A_{\mu}$. More specifically, given the symmetries adopted in
our set-up, the vector potential has only one non-zero component
$A_{\phi} \propto \max(q,0)$, where
\begin{align}
\begin{aligned}
q = 
    \begin{cases}
    \rho/\rho_{\rm max} - 0.2\,,& \text{for SANE}\\ \rho/\rho_{\rm
      max}\left({r}/{r_{\rm
        in}}\right)^3\sin^3\theta\exp\left(-{r}/{400}\right) - 0.01\,, &
    \text{for MAD}\,.
    \end{cases}
\end{aligned}
\end{align}

\begin{figure*}
    \centering \includegraphics[width=0.7\textwidth]{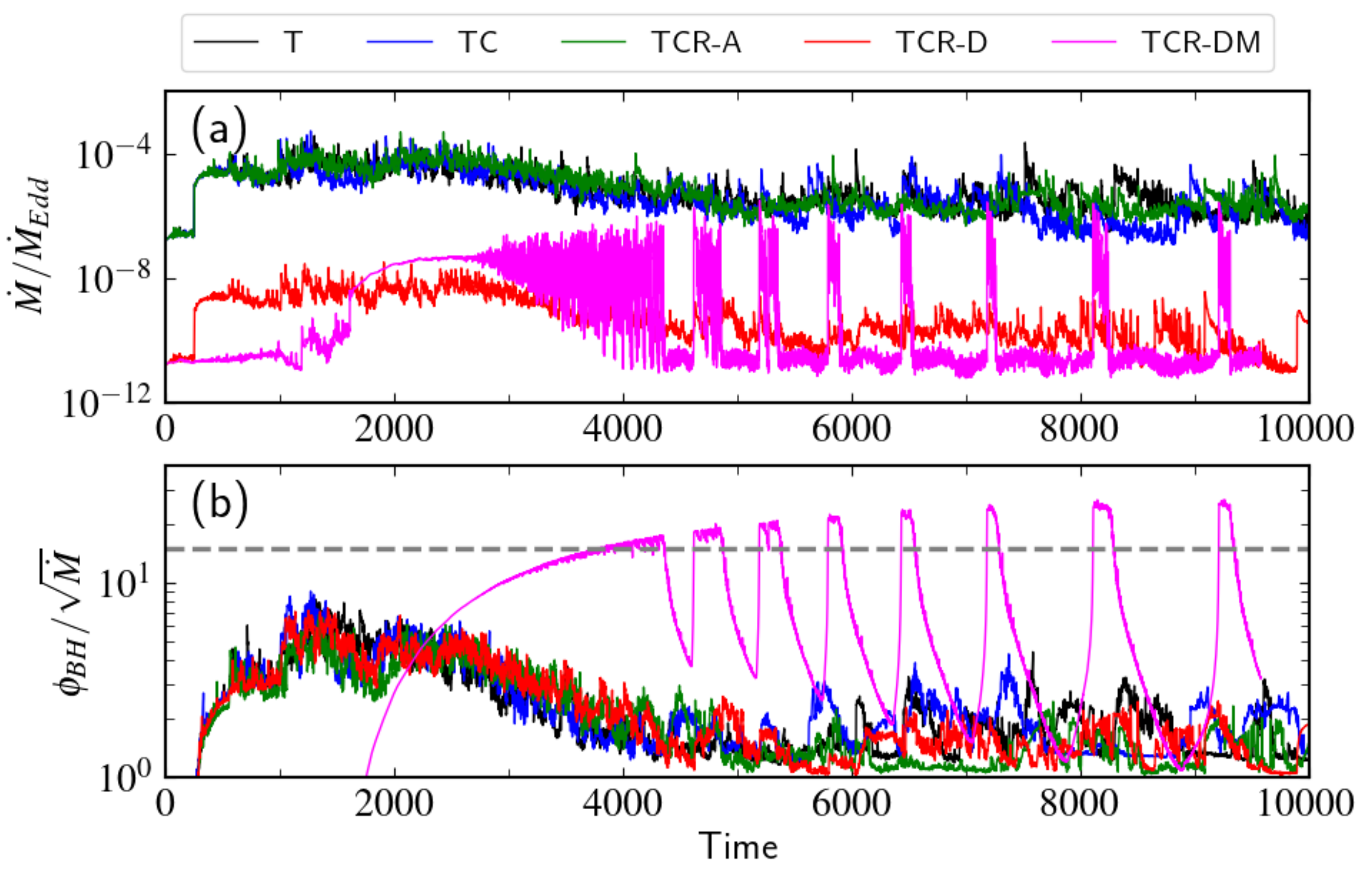}
    \caption{Evolution of the mass-accretion rate $\dot{M}/\dot{M}_{\rm Edd}$ as measured
      at the event horizon (top panel) and of
      the corresponding normalised magnetic flux $\phi_{\rm
        BH}/\sqrt{\dot{M}}$. The horizontal gray dashed line corresponds
      to $\phi_{\rm BH}/\sqrt{\dot{M}}=15$ and marks therefore the
      threshold for a MAD accretion mode.}
    \label{fig:acc-mag}
\end{figure*}

The strength of the initial magnetic field is determined by the specified
initial value of the maximum plasma-$\beta$ parameter, $\beta_{\rm
  max}=p_{\rm gas}/p_{\rm mag}\vert_{\rm max}$, which is specified at the
location of the rest-mass density maximum in the torus. For this study,
we consider $\beta_{\rm max}=100$ for all simulation models and perform
all the simulations in spherical modified Kerr-Schild coordinates in two
dimensions (axisymmetric) $(r,\theta)$ \citep{McKinney-Gammie2004}. The
simulation domain extends radially from well inside the black-hole event
horizon to $r=2500\,M_{\bullet}$ using a logarithmic spacing. In the
polar direction, we consider the domain ranges from a polar angle
$\theta=0$ to $\theta=\pi$ with uniform spacing. The numerical domain
is resolved with an effective grid resolution of $2048\times1024$ and
with three AMR levels. Furthermore, we set no-inflow boundary conditions
at the radial boundaries of the numerical domain. In contrast, we set the
scalar variables and the radial component of vectors to be symmetric at
the polar boundaries, while the azimuthal and polar vector components are
set to be antisymmetric at the polar boundaries.

In order to excite the magneto-rotational instability (MRI), we initially
apply a $1\%$ random perturbation to the gas pressure within the
torus. As it is customary in Eulerian, finite-volume codes, we adopt a
floor model to ensure that the GRMHD code can handle the low-density
regions, particularly close to the black hole and the rotation axis
\citep{Rezzolla-Zanotti2013}. More specifically, we set the floor
rest-mass density and pressure as $\rho_{\rm fl}=10^{-4} r^{-3/2}$ and
$p_{\rm fl}=(10^{-6}/3)r^{-5/2}$, respectively. In the case of the
electron pressure, we set $p_e=0.01p_{\rm fl}$ for $p_e\leq0.01p_{\rm
  fl}$ and $p_e=0.99p_{\rm g}$ for $p_e\geq0.99p_{\rm g}$. Note that we
recalculate the entropy of the gas $(\kappa_{\rm g})$ and of the
electrons $(\kappa_{\rm e})$ in the numerical cells where the flooring is
needed. In all simulations, we fix the mass and spin of the black hole to
be $M_{\bullet}=4.5\times10^6M_{\odot}$ and $a_*=0.9375$, respectively.

To obtain a rather broad and comprehensive investigation of the various
scenarios, we consider several simulation models, where details of these
models are summarised in Table \ref{tab-01}. More specifically, models
\texttt{T}, \texttt{TC}, and \texttt{TCR-A} are devised to study the
impacts of included physics in combination with turbulent heating
(\texttt{T}), Coulomb interaction (\texttt{C}), and radiative cooling
(\texttt{R}). Furthermore, in order to understand the role played by the
accretion rate in the thermodynamics of electrons, we consider models
\texttt{TCR-A}, \texttt{B}, \texttt{C}, \texttt{D}, and \texttt{E} with
different values of the mass-accretion rate, namely $\dot{M} = 10^{-4} -
10^{-11} \dot{M}_{\rm Edd}$, where $\dot{M}_{\rm
  Edd}=1.4\times10^{17}M_{\bullet}/M_{\rm \odot}$ is the Eddington
mass-accretion rate. Finally, model \texttt{TCR-DM} is designed to study
the difference in the electron thermodynamics between the SANE (model
\texttt{TRC-D}) and the MAD accretion flows. Before moving to the
  results, we recall that GRMHD simulations not involving
  radiative-transfer processes are independent of the black-hole mass,
  which can be therefore simply rescaled. On the other hand, when
  radiative-transfer is taken into account, the results do depend on the
  mass of the black hole since the radiative processes depend on the
  mass-accretion rate. In view of this, in our simulations we keep the
  black-hole mass fixed and vary only the mass-accretion rate.

\section{Results}
\label{sec:three}

To illustrate the temporal properties of the accretion flow, we show in
Fig.~\ref{fig:acc-mag} the profile of the mass-accretion rate
($\dot{M}/\dot{M}_{\rm Edd}$) and of the magnetic-flux rate at the 
black hole horizon $(\Phi_{\rm BH}/\sqrt{\dot{M}})$ for different
simulation models: \texttt{T} (black), \texttt{TC} (blue), \texttt{TCR-A} 
(green), \texttt{TCR-D} (rad),
\texttt{TCR-DM} (magenta). The grey line in the Fig.~\ref{fig:acc-mag}b
corresponds to $\Phi_{\rm BH}/\sqrt{\dot{M}}=15$, which is
commonly adopted as the threshold in the magnetic-flux rate
distinguishing SANE accretion ($\Phi_{\rm BH}/\sqrt{\dot{M}}<15$) 
from MAD accretion modes ($\Phi_{\rm
  BH}/\sqrt{\dot{M}}>15$; details on how these quantities are
calculated have been reported by \cite{Porth-etal2017}.

As it is well-known in these simulations, the weak poloidal magnetic
field lines wind up as the evolution proceeds and generate the toroidal
component. The differential rotation in the torus triggers the MRI, which
helps in the outward transport of angular momentum, triggering and
regulating the accretion process. Further, the MRI also amplifies the
toroidal component of the magnetic field
\citep[e.g.,][]{Begelman-Pringle2007}. We set the effective resolution
such that the fastest growing MRI mode is resolved with a quality factor
$Q_\theta\gtrsim6$, except for the very high-density region for the SANE
modes \citep[see, e.g., Fig. 13 of][for a representative
  example]{Nathanail-etal2020}, which is required to ascertain that MRI
is active in our simulations \citep[e.g.,][]{Sano-etal2004}. Once the
accretion process is set, we observe that the mass-accretion rates for
all the simulation models (\texttt{T}, \texttt{TC}, \texttt{TCR-A},
\texttt{TCR-D}) reach a quasi-steady-state after a certain time $t
\gtrsim 1500\,M_{\bullet}$, except for model \texttt{TCR-DM}, which
reaches a quasi-steady-state after $t \gtrsim 2500\,M_{\bullet}$ as a
result of the larger inner radius of the torus $r_{\rm in}$ (see
Fig.~\ref{fig:acc-mag}a).

\begin{figure*}
    \centering
    \includegraphics[width=1.0\textwidth]{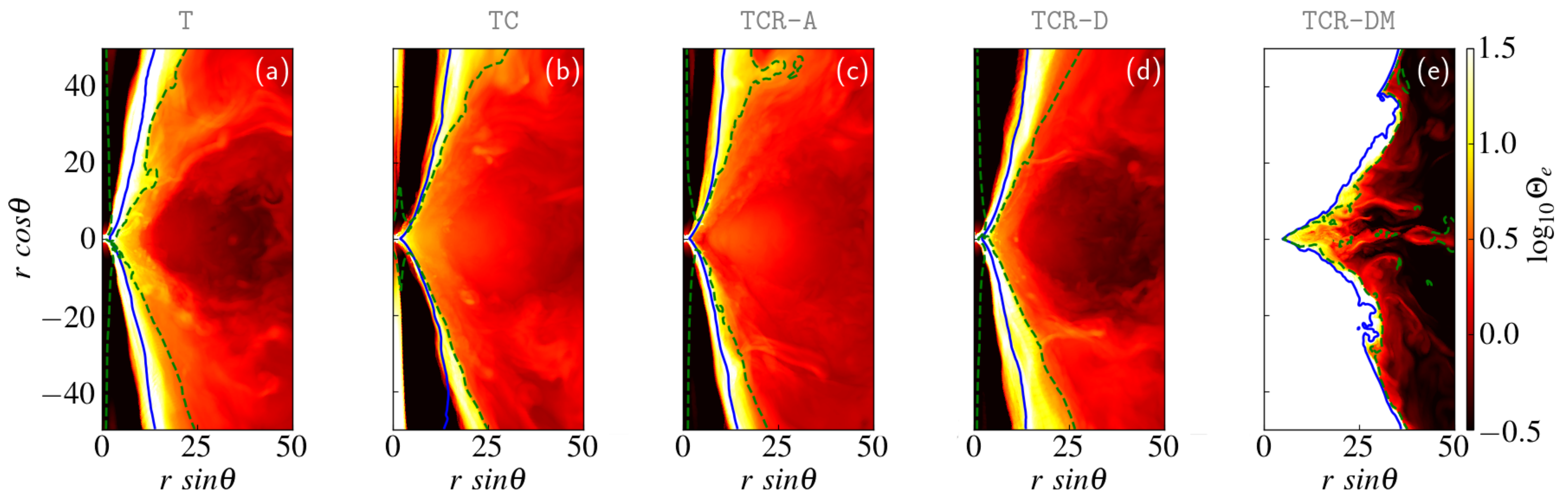}
    \caption{Spatial distributions of the dimensionless electron
      temperatures $\Theta_e$ for the different simulation models:
      \texttt{T}, \texttt{TC}, \texttt{TCR-A}, \texttt{TCR-D}, and
      \texttt{TCR-DM} [panels (a), (b), (c), (d), and (e),
        respectively]. The distributions are time-averaged within the
      window $t=8000-9000\,M_{\bullet}$, and the blue solid and green
      dashed lines mark the boundaries where $\sigma=1$ and the Bernoulli
      parameter $-hu_t=1$, respectively.}
    \label{fig:ele-temp}
\end{figure*}

After $t \gtrsim 4000\,M_{\bullet}$, the mass-accretion rate and the
magnetic-flux rates for model TCR-DM show the large oscillations that are
typical of the MAD accretion mode and that reflect the periodic quenching
of the accretion flow close to the black hole
\citep[e.g.,][]{Tchekhovskoy-etal2011,Dihingia-etal2021}. Also, in
Fig.~\ref{fig:acc-mag}b, we observe that with time, $\Phi_{\rm
  BH}/\sqrt{\dot{M}}$ increases and reaches a maximum
value. Except for model TCR-DM (magenta), the value of $\Phi_{\rm
  BH}/\sqrt{\dot{M}}$ drops at time $t \sim 1500\,M_{\bullet}$
and saturates after $t \gtrsim 5000\,M_{\bullet}$, with $\Phi_{\rm
  BH}/\sqrt{\dot{M}} \sim 1.5$. The profiles of $\Phi_{\rm BH}/\sqrt{\dot{M}}$
essentially suggest that all these simulation models are in the SANE
state. The only exception is for model TCR-DM, where the magnetic-flux
rate at the black hole horizon increases monotonically and exceeds the
saturation limit for a MAD configuration after
$t\gtrsim4000\,M_{\bullet}$, i.e., $\Phi_{\rm BH}/\sqrt{\dot{M}} \sim 15$
\citep[e.g.,][]{Tchekhovskoy-etal2011,Mizuno-etal2021}. We recall that in
a MAD accretion mode, the inner part of the accretion flow releases a
substantial amount of the accumulated magnetic flux through magnetic
eruption events \citep[e.g.,][]{Igumenshchev2008,
  Dexter-etal2020,Porth-etal2021}. When this happens, the magnetic flux
drops to a smaller value $(\Phi_{\rm BH}/\sqrt{\dot{M}} \ll
15)$ up until sufficient magnetic flux is advected by the flow and
accumulates near the horizon, leading to a MAD cycle with an increase in
the magnetic-flux rate. This process continues in the form of
quasi-period oscillations throughout the simulation time.
Note that in more realistic three-dimensional (3D) GRMHD simulations, 
the eruption events in MAD configuration are not as abrupt as seen in 
Fig. \ref{fig:acc-mag}. The profile for $\Phi_{\rm BH}/\sqrt{\dot{M}}$ 
for the MAD regime in 3D GRMHD simulations is much more laminar compared 
to asymmetric models with episodic eruption events due to the presence of 
interchange instabilities in the azimuthal direction 
\citep[e.g., see][]{Mizuno-etal2021,Dexter-etal2020,Porth-etal2021}.

During the quasi-steady-state, we observe various phenomena driven by
turbulence, reconnection events, plasmoids, and so on. These features are
qualitatively similar to the studies done in the single-fluid ideal GRMHD
framework \citep[e.g., ][]{Porth-etal2017, Porth-etal2019,
  Nathanail-etal2020, Dihingia-etal2021, Porth-etal2021}. However, since
the main interest in our investigation is the study of the time-averaged
properties of the accretion flow, we will not discuss in detail the
properties of the plasma -- which can be found well described in the
references above -- and concentrate instead on a number of time-averaged
flow quantities. We should also note that our simulations treat the
accretion flow only at low accretion rates, which remains optically thin.
Under these conditions, the radiative-cooling processes are not strong
enough to collapse the geometrically thick torus to a thin-disc
structure. 

\subsection{Distribution of the temperatures}

\begin{figure*}
    \centering
    \includegraphics[width=1.0\textwidth]{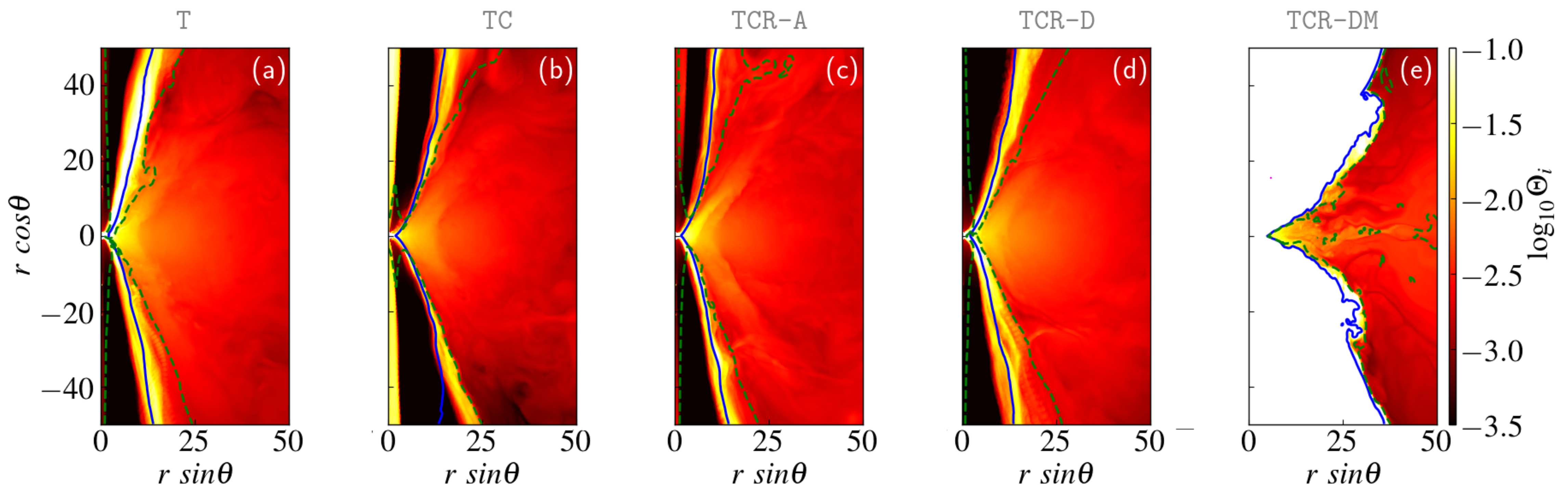}
    \caption{Same as in Fig.\ref{fig:ele-temp} but for the time-averaged
      dimensionless ion temperature $\Theta_i$}.
    \label{fig:prot-temp}
\end{figure*}

In Figs. \ref{fig:ele-temp} and \ref{fig:prot-temp}, we show the
time-averaged and dimensionless electron temperature $\Theta_e$ and the
dimensionless ion temperature $\Theta_i:=k_{\rm B}T_i/m_p c^2$ for the
simulation models \texttt{T}, \texttt{TC}, \texttt{TCR-A},
\texttt{TCR-D}, and \texttt{TCR-DM} (the names of various models are
reported at the top of each panel), within a temporal domain
$t=8000-9000\,M_{\bullet}$. Each panel shows the distribution in a small
region near the black hole, with the blue-solid and the green-dashed
lines indicating the contours with magnetisation $\sigma=1$ and a
Bernoulli parameter $-hu_t=1$, respectively, where $h$ is the
specific enthalpy of the flow. The region around the
rotation axis of the black hole with $\sigma>1$ and $-hu_t>1$ is known as
the funnel region of the flow. 

The region surrounding the polar funnel contributes to the wind from the
disc, which is also known as the jet-sheath region of the flow. In
Fig.~\ref{fig:ele-temp}a (model \texttt{T}), electrons receive a fraction
of the energy gained by the flow via the turbulent heating. As a result,
we observe a higher electron temperature in the region close to the black
hole and in the jet-sheath region. We note that such an efficient
electron heating around the shearing region between bound and unbound
material $(-hu_t\sim1$, green dashed contour) was reported by also by
several previous studies \citep[e.g.,][]{Chael-etal2019,
  Dexter-etal2020a, Anantua-etal2020, Mizuno-etal2021}.

With the inclusion of the Coulomb-interaction term (panel
\ref{fig:ele-temp}b, model \texttt{TC}), the electrons gain an extra
fraction of energy from the ion fluid. Consequently, the electron
temperature becomes even higher for model TC throughout the domain as
compared to model \texttt{T}. If the electrons are allowed to cool
through the radiation-cooling processes (model \texttt{TCR-A}, panel (c)
in Fig.~\ref{fig:ele-temp}), the electron temperature then decreases in
the region close to the black hole (see the dark red colour) when
compared to models \texttt{TC} and \texttt{T}. Furthermore, when the
mass-accretion rate is set to be smaller (model \texttt{TCR-D}, panel (d)
in Fig.~\ref{fig:ele-temp}), the overall density of the flow and the
magnetic-field strength decrease as compared to the higher accretion rate
cases. Consequently, the radiative cooling due to the Bremsstrahlung and
the synchrotron processes becomes less efficient. On the contrary, the
turbulent heating process remains unaffected as it depends only on the
plasma-$\beta$ and the ratio of temperatures. These quantities suffer
minimum changes due to the decrease in accretion rate. Therefore, the
electrons remain hot close to the black hole and in the jet-sheath
region. Also, together with the decrease of the accretion rate, the
Coulomb-interaction rate decreases, so that the electrons in the
high-density region are comparatively cooler when compared to the high
accretion flow (model \texttt{TCR-A}, panel (c) in
Fig.~\ref{fig:ele-temp}).

The effects of the radiative cooling and of the heating in the SANE and
MAD accretion modes can be appreciated by comparing panels (d) and (e) in
Fig.~\ref{fig:ele-temp}. More specifically, in the MAD mode (model
\texttt{TCR-DM}), the turbulent heating dominates in the polar region,
resulting in a very hot funnel with temperatures that are considerably
higher, i.e., more than a factor of $\sim 10-50$, than in the SANE mode
(model \texttt{TCR-D}). However, in the high-density regions, the
electrons are actually colder in the MAD mode and by a factor $\sim
3-5$. Note that in both cases the electrons remain hot close to the
black hole, since the Coulomb-interaction term dominates over the
radiative cooling in these regions.

\begin{figure*}
    \centering
    \includegraphics[width=1.0\textwidth]{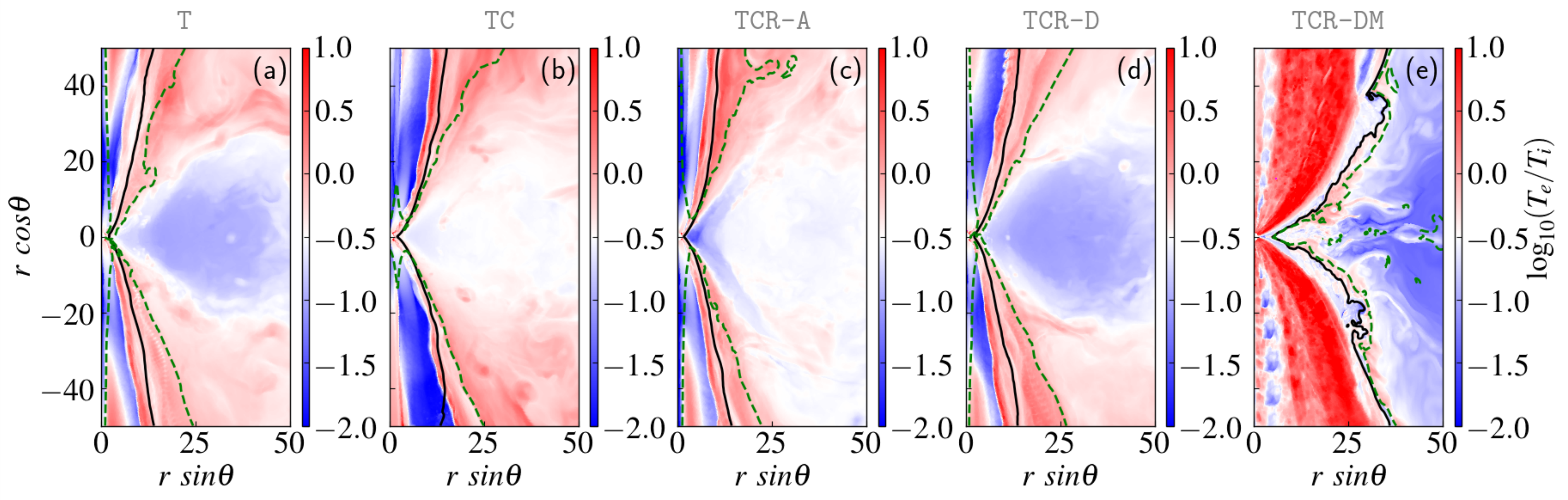}
    \caption{Ratio of the time-averaged electron-to-ion temperatures
      $T_e/T_i$ (see Figs.~\ref{fig:ele-temp} and
      \ref{fig:prot-temp}). The black solid and green dashed lines mark 
      the boundaries where $\sigma=1$ and the Bernoulli
      parameter $-hu_t=1$, respectively.}
    \label{fig:tetp}
\end{figure*}

Figure \ref{fig:prot-temp} reports the distribution of the ion
temperature for the different models considered and effectively shows the
impact of the Coulomb-interaction term on this quantity. Note that the
Coulomb interaction actually acts as a cooling term for the ions. As a
result, with its inclusion, the ions close to the black hole and the
jet-sheath region are significantly colder (compare panels (a) and (b) of
Fig.~\ref{fig:prot-temp}). Since the electrons are allowed to cool
through radiative cooling, the difference between the electron and ion
temperatures increases, and as a result, the efficiency of the Coulomb
interaction also increases (see Eq. \ref{eq:cc}), leading to a cooling of
the ions in the high-density region (see panel (c) in
Fig.~\ref{fig:prot-temp}). With the decrease of the accretion rate (see
panel (d) in Fig.~\ref{fig:prot-temp}), the efficiency of both the
radiative cooling and the Coulomb interaction drops. As a result, the
ions in the high-density region remain hotter than in the highly
accreting cases (Fig.~\ref{fig:prot-temp}c). Furthermore, by comparing
panels (d) and (e) of Fig.~\ref{fig:prot-temp}, it is possible to note
that the ions in the funnel region are hotter for the MAD mode than in
the SANE mode for the same accretion rate. At the same time, the ions
in the high-density region are colder in the MAD accretion mode than in
the SANE accretion mode.

To better understand the contribution of the heating and cooling terms,
we report in Fig.~\ref{fig:tetp} the distribution of the temperature
ration $T_e/T_i$ averaged over a time window of
$t=8000-9000\,M_{\bullet}$ and for all of the different models considered
in Figs. \ref{fig:ele-temp} and \ref{fig:prot-temp}. We note that for the
model with only turbulent heating (model \texttt{T},
Fig.~\ref{fig:tetp}a), we observe that the heating of the electrons is
particularly strong in the jet-sheath region, where $-hu_t \sim 1$ and
$T_e/T_i \sim 1-10$. However, in the funnel region, the electrons remain
cold, i.e., $T_e/T_i \lesssim 0.01$. With the inclusion of the
Coulomb-interaction term, the electrons heat up and consequently, the
ratio $T_e/T_i$ increases throughout the domain, except in the funnel
region (see panel (b) in Fig.~\ref{fig:tetp}); this is because the
electron-ion collision rate is suppressed in these regions as a result
of the very low density, so that ions cannot transfer their energy to
the electrons.

\begin{figure}
    \centering
    \includegraphics[scale=0.43]{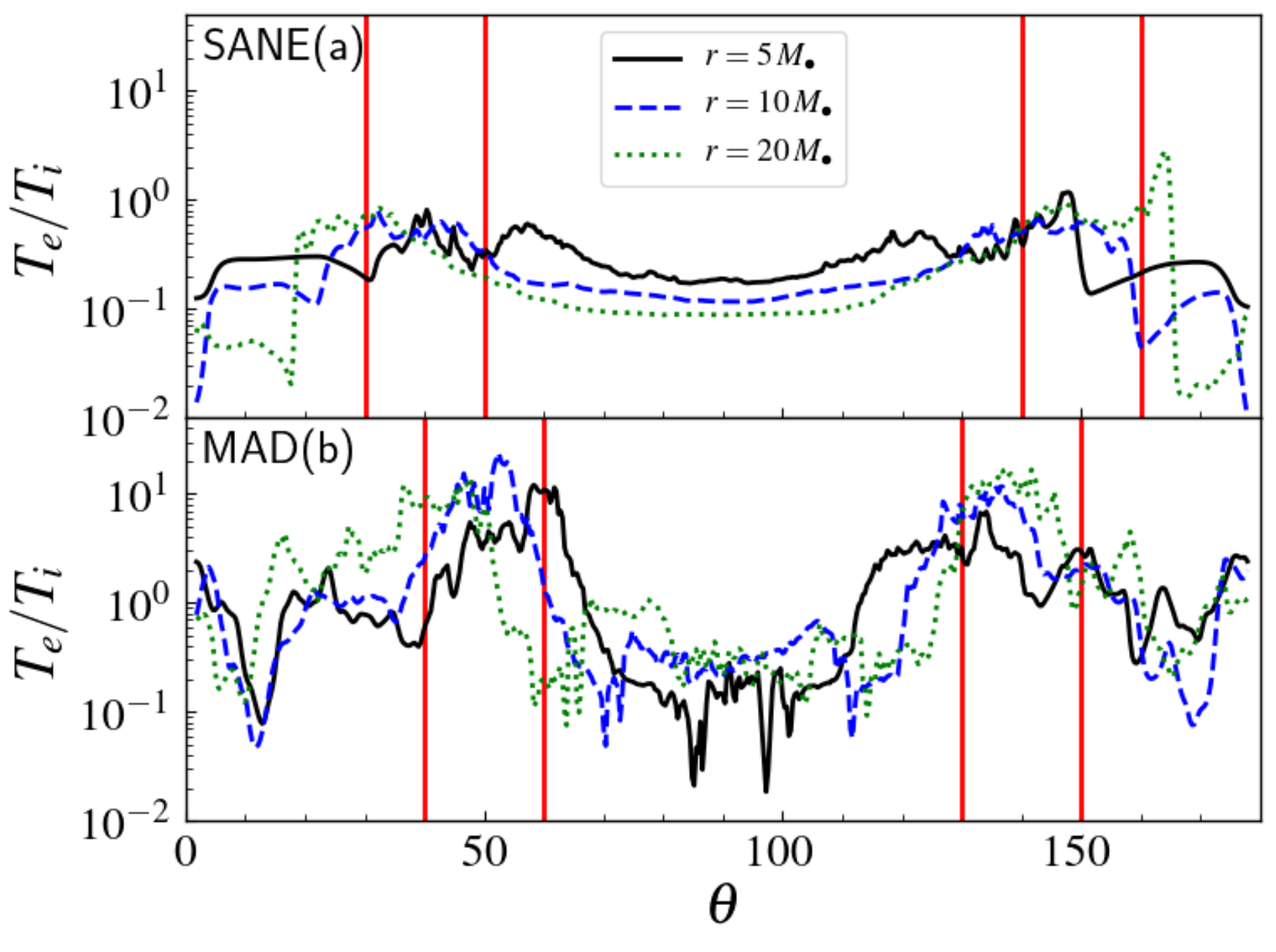}
    \caption{Plots of temperature ratio $T_e/T_i$ shown as a function of
      the polar angle $\theta$ (in degrees) at different radii
      $r=5,10,20\,M_{\bullet}$ for models \texttt{TCR-D} (top panel) and
      \texttt{TCR-DM} (bottom panel). The two vertical red lines on both
      sides of the equatorial plane ($\theta=90$) roughly denote the
      jet-sheath region; regions the left or to the right of these two
      lines correspond to the funnel.}
    \label{fig:tetpline}
\end{figure}

We note that when the radiative cooling is included (panels (c) and (d)
in Fig.~\ref{fig:tetp}), it overpowers the heating processes close to the
black hole, resulting in a lower value of $T_e/T_i$ there. When comparing
the SANE and MAD accretion modes (panels (d) and (e) in
Fig.~\ref{fig:tetp}), it is possible to note that the distribution of
$T_e/T_i$ is different in the jet-sheath and in the funnel regions. More
specifically, in the MAD mode, both the electron and ion temperatures
around the equatorial plane and far from the black hole are smaller than
those encountered in the SANE mode (see Figs. \ref{fig:ele-temp} and
\ref{fig:prot-temp}). On the contrary, near the black hole, the behaviour
is the opposite. To understand the origin of this difference, we report
in Fig.~\ref{fig:tetpline} the ratio $T_e/T_i$ as a function of the polar
angle ($\theta$) at different radii close to the black hole for both the
SANE (left panel) and the MAD (right panel) simulations. The two panels
essentially suggest that for the SANE modes, the ratio $T_e/T_i \lesssim
1$ in the jet-sheath and the ratio $T_e/T_i\lesssim 0.1$ in the funnel
region. However, for the MAD mode, the ratio $T_e/T_i \gtrsim 1$ in both
the jet-sheath and funnel. Earlier studies with turbulent heating also
suggest that a hot funnel flow develops in the case of MAD accretion
modes \citep{Dexter-etal2020a, Mizuno-etal2021}. We should note
  that we consider the SANE models to have a magnetic flux $\sim 10$
  times smaller than the MAD model (see Fig. \ref{fig:acc-mag}).  In
  principle, we can raise the magnetic flux accumulation around the event
  horizon of a black hole by setting the initial condition of the
  accretion flows such as torus size and initial magnetic field strength.
  If the normalised magnetic flux is still within the SANE limit (e.g.,
  $\dot{\Phi}/\sqrt{\dot{M}} \sim 5-10$), the accretion flow is still
  referred to as SANE. However,the temperature profiles of electrons and
  ions may not be as distinct as stated in this study.

\begin{table}
\centering
  \begin{tabular}{| l  |c | c | c | c | c |}
    \hline
    Model  & $R_{1,p}$ & $R_{2,p}$ & $R_{3,p}$ & $\beta_{\rm br}$\\ 
    \hline
    \texttt{T}      & 2.55 & 1.36 & 8.52  & 49.7 \\
    \texttt{TC}     & 2.37 & 0.89 & 2.72  & 8.24 \\
    \texttt{TCR-A}  & 2.05 & 9.49 & -5.54 & 10.1 \\
    \texttt{TCR-B}  & 2.56 & 3.24 & 3.60  & 18.1 \\
    \texttt{TCR-C}  & 2.78 & 1.94 & 6.87  & 38.4 \\
    \texttt{TCR-D}  & 2.80 & 1.68 & 8.59  & 55.0 \\
    \texttt{TCR-DM} & 4.50 & 2.26 & 22.9  & 517  \\
    \texttt{TCR-E}  & 2.81 & 1.52 & 9.31  & 55.8 \\
    \hline
  \end{tabular}
\caption{Best-fit parameters $R_{1,p}, R_{2,p}, R_{3,p}$, and $\beta_{\rm
    br}$ for the different simulation models. These values are obtained
  from the time-averaged temperature ratios $T_e/T_i$ over a time window
  $t=8000-9000\,M_{\bullet}$. }
\label{tab-02}
\end{table}

\begin{figure*}
    \centering
    \includegraphics[scale=0.4]{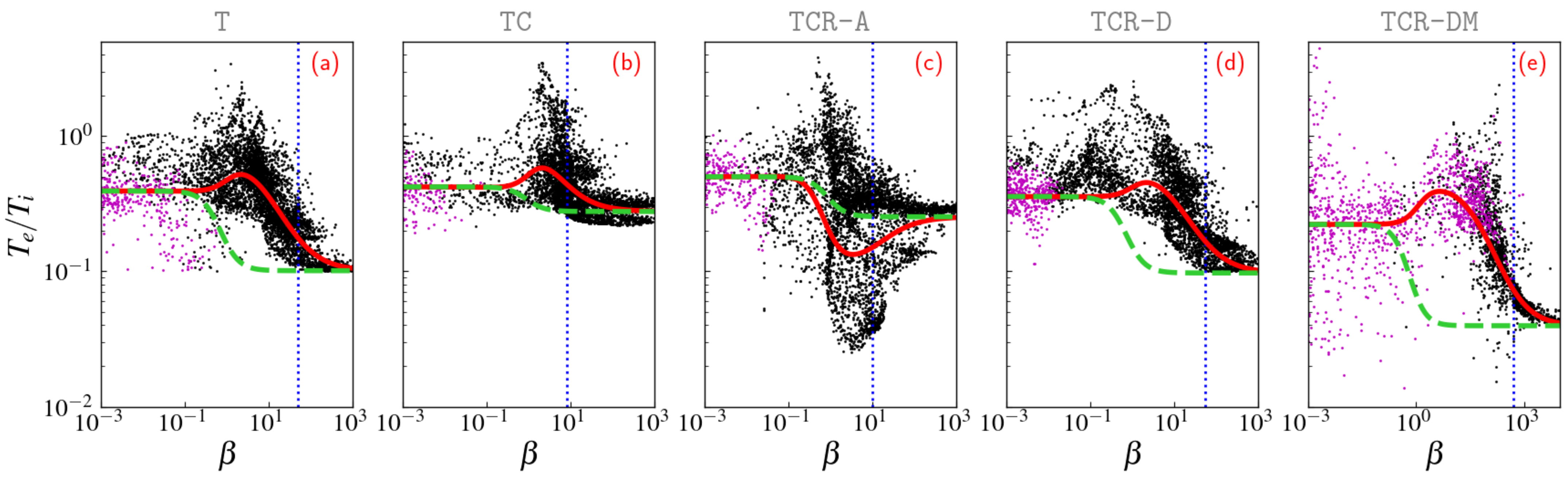}
    \includegraphics[scale=0.4]{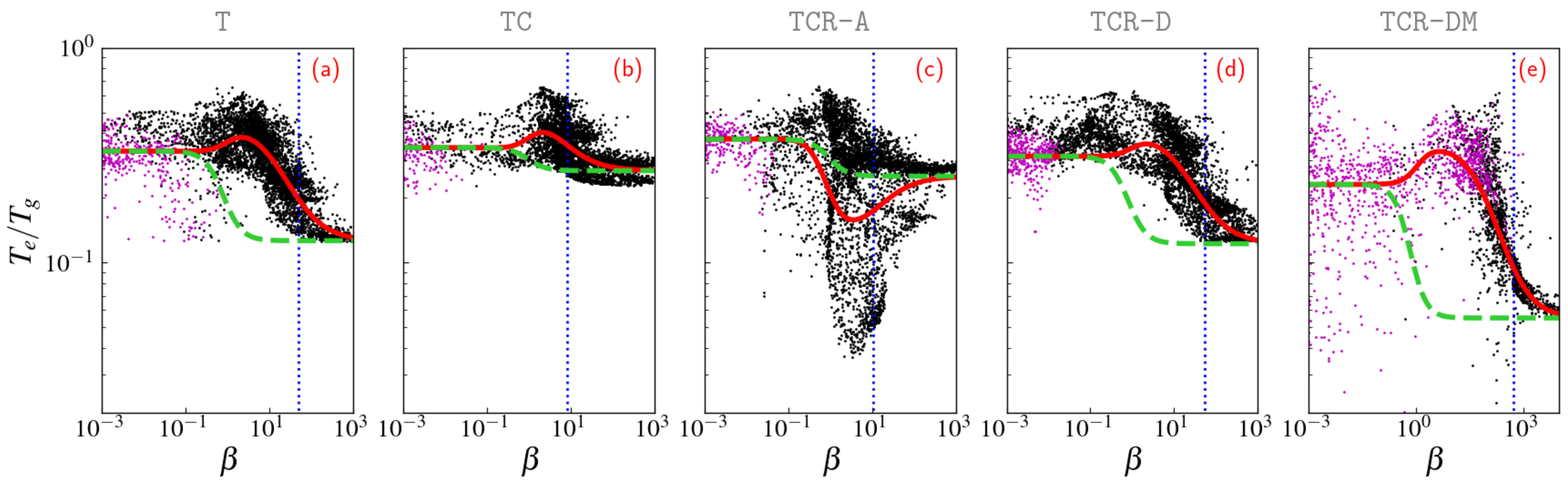}
    \caption{Distributions of the temperature ratio $T_e/T_i$ (top
      panels) and of $T_e/T_g$ (bottom panels) as a function of
      plasma-$\beta$ for the simulation models \texttt{TCR-A},
      \texttt{TCR-D}, and \texttt{TCR-DM} [panels (a), (b), and (c),
        respectively]. Each dot refers to a cell in the computational
      domain, with the black dots referring to the region with $\sigma<1$
      (roughly representative of the disc) and the magenta dots to the
      region with $\sigma>1$ (roughly representative of the jet). The
      solid red curves corresponds to our new $R-\beta$ relations, while
      the dashed green lines correspond to the usual $R-\beta$
      relation. The vertical dotted blue lines mark the value of
      $\beta=\beta_{\rm br}$.}
    \label{fig:corre}
\end{figure*}

\subsection{Correlation of the temperatures}

In order to fully understand the correlation between the temperatures, we
collect the ratio of the temperatures $T_e/T_i$ and $T_e/T_g$ from all
the grid-cells in the computational domain with
$r<100\,M_{\bullet}$. The distribution of the corresponding temperature
ratios $T_e/T_i$ and $T_e/T_g$ as a function of the plasma-$\beta$ is
shown in the top and bottom panels of Fig.~\ref{fig:corre},
respectively. In all of the panels of Fig.~\ref{fig:corre}, each dot
represents the value measured in a given grid-cell over a time window
$t=8000-9000\,M_{\bullet}$. The black dots are collected from the parts
of the domain with $\sigma<1$ and thus corresponding to the ``disc
region'', while the magenta dots are from cells with $\sigma>1$ and thus
can be considered the ``jet region''. Note that how in the case of a SANE
accretion mode, i.e., for panels (a)--(d) in Fig.~\ref{fig:corre}, these
magenta values are mostly restricted to small values of the
plasma-$\beta$ parameter; on the other hand, for a MAD accretion mode,
i.e., for panels (e), they can also be associated with values of $\beta
\lesssim 10^2$, although they tend to be absent in the regions with
$\beta \gg 1$.

Note how the different scatterings of the dots clearly show the impact of
various mechanisms on the temperatures. In particular, the inclusion of
the Coulomb-interaction term heats-up the electrons quite irrespective of
the plasma-$\beta$ (see panels (a) and (b) of Fig.~\ref{fig:corre}). As a
result, the value of $T_e/T_i$ increases in the whole range of
plasma-$\beta$ for model \texttt{TC} when compared to model
\texttt{T}. The increase in the temperature ratios is prominent for the
higher plasma-$\beta$ range (i.e., $\beta>10$), as the Coulomb
interaction is more efficient in the high-density region of the disc. As
we include the radiative-cooling processes (panel (c)), electrons are
cooled and lower values of $T_e/T_i$ and $T_e/T_g$ are reached for $\beta
\sim1$. However, in the region with low plasma-$\beta$, the turbulent heating
remains stronger than the radiation-cooling term and the electrons in
these regions remain comparatively hot. Note the appearance of a local
minimum in the temperature ratios in panels (c) of Fig.~\ref{fig:corre},
which indicates that the cooling is particularly important and actually
dominating in the range $\beta\sim1-10$.

With the decrease in the accretion rate, the efficiency of the Coulomb
heating and of the radiative cooling also decreases (panels (d) and (e)
of Fig.~\ref{fig:corre}). On the contrary, the rate of turbulent heating
is independent of the accretion rate (see Eq.~\eqref{eq:05}). As a
result, electrons with lower plasma-$\beta$ remain hot while electrons
with higher plasma-$\beta$ remain cold when the accretion rates are
small. Interestingly, the ratio $T_e/T_i$ becomes greater than unity in
the regions where the turbulent heating of electrons is efficient and the
radiative cooling is less efficient. These regions appear around the
funnel (see Fig.~\ref{fig:tetp}). However, the gas temperature always
remains higher than that of the electrons throughout the simulation
domain (see bottom panels of Fig.~\ref{fig:corre}).

Figure \ref{fig:tetp} suggests that the ratio $T_g/T_e$ or
$T_i/T_e$ shows a broad correlation with the plasma-$\beta$
parameters. Moreover, the trend of the temperature ratio is quite similar
for SANE as well as MAD modes (panels (d) and (e) of
Fig.~\ref{fig:tetp}). Therefore, it is possible to relate the
temperatures ratios $T_g/T_e$ and $T_i/T_e$ with the properties of the
plasma-$\beta$ in the accretion flow. We can do this by fitting the data
in Fig \ref{fig:corre} with a simple function
of plasma-$\beta$ parameter that can recover the widely used $R-\beta$
prescription in the suitable limits. In particular, we adopt the new
ansatz
\begin{align}
    \frac{T_k}{T_e}= \frac{1}{1+\beta^2}R_{1,k} +
    \frac{\beta^2}{1+\beta^2} R_{2,k} + \frac{\beta}{\beta+\beta_{{\rm
          br}}} R_{3,k} \,.
    \label{eq:rbeta}
\end{align}
where $k=g, i$ and $\beta_{{\rm br},k}$ is what we refer to as the ``the
plasma-$\beta$ break'', i.e., a characteristic value of the $\beta$
parameter where the temperature ratio undergoes a sharp change (see
below). Two important remarks are needed at this point. First,
Eqs.~\eqref{eq:rbeta} seem to imply that there are six new coefficients
$R_{1-3,g}$ and $R_{1-3,p}$; in practice, however, only three are
linearly independent since the other three can be derived through
relations that are simple to derive in the case of an ideal-fluid
equation of state, namely
\begin{align}
R_{i,g} & = \frac{\tilde{\Gamma}_g - 1}{\tilde{\Gamma}_i -1}\left(R_{i,p}
+ \frac{\tilde{\Gamma}_i - 1}{\tilde{\Gamma}_e -1}\right)\,, \qquad\qquad i=1,2\\
R_{3,g}& = \left(\frac{\tilde{\Gamma}_g -1}{\tilde{\Gamma}_i - 1}\right) R_{3,p}\,.
\label{eq:r-rel}
\end{align}
Second, note that the standard $R-\beta$ relation can be obtained by
setting $R_{1,p}=R_{\rm low}$, $R_{2,p}=R_{\rm high}$ and $R_3=0$, i.e.,
\begin{align}
    \frac{T_k}{T_e}= \frac{R_{\rm low}}{1+\beta^2} + \frac{R_{\rm high}
      \beta^2}{1+\beta^2}\,.
    \label{eq:rbetaorg}
\end{align}

\begin{figure*}
    \centering
    \includegraphics[scale=0.55]{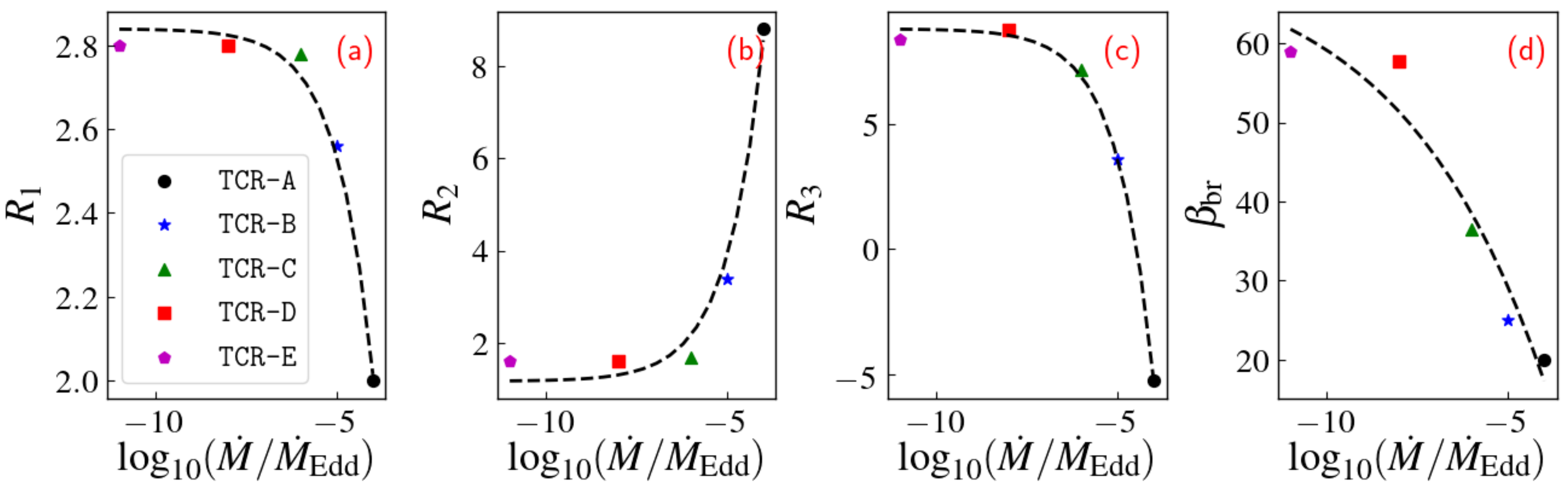}
    \caption{Values of the best-fit parameters $R_{1-3,p}$ and of
      $\beta_{\rm br}$ as a function of the accretion rate expressed in
      CGS units and normalised to the Eddington accretion rate. Different
      symbols are used to refer to the different models considered, while
      black dashed lines correspond to the fitting functions of the
      data.}
    \label{fig:para}
\end{figure*}

The new functional form of the temperature ratios \eqref{eq:rbeta} is
built on the expectation that the electrons should be cooled by the
thermal synchrotron emission in the moderately magnetised region with
$\beta \sim 1$. In addition, the electrons should have a higher
temperature in the highly magnetised regions as a result of the strong
turbulent heating. Under these considerations, and for $\beta \ll 1$, the
temperature ratios reduce to $T_i/T_e \simeq R_1$ and $T_g/T_e \simeq
R_1$. For larger values of $\beta$, and depending on the heating and
cooling processes, the ratio of the temperatures may increase or
decrease. However, for $\beta \gg \beta_{\rm cr} \gg 1$, the temperature
ratios tend to reach constant values $T_i/T_e \simeq ~T_g/T_e \simeq
R_{2,p} + R_{3,p}$, so that the standard value of $R_{\rm high}$ normally
employed in the standard $R-\beta$ relation \eqref{eq:rbetaorg} is
actually given by $R_{\rm high} = R_{2,p} + R_{3,p}$.

In the various panels of Fig.~\ref{fig:corre} we show with a solid red
line the new $R-\beta$ relations \eqref{eq:rbeta} as obtained from the
best-fit parameters for the various models considered; for the ratio
$T_i/T_e$ these best-fit parameters $R_{1-3,p}$ are reported in Table
\ref{tab-02} for the various models considered, while for the $T_g/T_e$
ratio they can be obtained by using expressions \eqref{eq:r-rel}. In all
of the panels of Fig.~\ref{fig:corre} we also show the value of
$\beta_{\rm br}$ with a vertical blue dotted line. Note that the new
$R-\beta$ relation broadly follows the trend of the ratio of the
temperatures as a function of the plasma-$\beta$ of the fluid. For
comparison, in Fig.~\ref{fig:corre}, we also overlay the standard
$R-\beta$ function with $R_{\rm low}=R_1$ and $R_{\rm high} = R_2 + R_3$
with a green dashed line. As expected, the standard $R-\beta$ relation
shows a good agreement in the range $\beta\lesssim 0.1$ and $\beta\gtrsim
100$ for all of the models considered, in the range of $0.1
\lesssim\beta\lesssim100$, it does not provide an accurate description of
the thermodynamics of the accreting flow. We note that these three
different regions in the plasma-$\beta$ correspond to spatially different
regions, namely, the funnel/shearing region ($\beta\lesssim 0.1$), the
jet-sheath region ($0.1\lesssim\beta\lesssim100$) and the high-density
disc region ($\beta\gtrsim 100$). As a result, the distributions
illustrated in Fig.~\ref{fig:corre} clearly indicate that the standard
$R-\beta$ relation does not provide a faithful description of the
electron temperature in the jet-sheath region and that it tends to
systematically underestimate it; the only exception is for model
\texttt{TCR-A}, where a slight overestimate is present in the cooling
dominated regions.

Before concluding this section, we note that it is interesting to
investigate if the new values of the $R_{1-3,p}$ coefficients provided in
Table~\ref{tab-02} show some correlation with other bulk properties of
the flow, most importantly with the mass-accretion rate. To this scope,
we report in Fig.~\ref{fig:para}, the values of $R_{1-3,p}$ and of
$\beta_{\rm br}$ as a function of the accretion rate normalised to the
Eddington accretion rate. Each point in the figure refers to a specific
simulation, as indicated in the legend, and the reported values of the
$R_{1-3,p}$ parameters represent the best-fits of the distributions of
the temperature ratios reported in Fig.~\ref{fig:corre} and which, we
recall, are time-averaged over the window $t = 8000-9000\,M_{\bullet}$. A
brief inspection of Fig.~\ref{fig:para} clearly shows that the best-fit
parameters correlate very well with the mass-accretion rate, and in
principle, analytic fitting functions can be computed (see black dashed
lines in Fig.~\ref{fig:para}). The explicit forms of the fitting
functions are reported in Appendix A. These functions are important to
show the correlation of the $R_{1-3,p}$ or $\beta_{\rm br}$ with the
accretion rate. However, one must note that we obtained these functions
by computing best-fit parameters and accumulated errors may be
significant during this process. Nonetheless, these functions provide
ranges of these parameters ($R_{1-3,p}$, $\beta_{\rm br}$) at different
accretion rates for SANE models.

Note that the values of the parameters change drastically as the
accretion rate is higher than a certain value, namely,
$\dot{M}\gtrsim10^{-7} \dot{M}_{\rm Edd}$. This behaviour suggests that
the contribution of the heating and cooling to the thermodynamics of
electrons and ions becomes significant for accretion rates
$\dot{M}\gtrsim10^{-7} \dot{M}_{\rm Edd}$. More importantly, since the
standard coefficient $R_{\rm high}$ is simply the sum of $R_{2,p}$ and
$R_{3,p}$, Fig.~\ref{fig:para} highlights that $R_{\rm high}$ is not
independent of the accretion rate and should not be chosen arbitrarily.

\section{Conclusions \& discussion}
\label{sec:four}

We have employed a self-consistent approach to study magnetised and
radiatively cooled two-temperature accretion flows around a Kerr black
hole in two spatial dimensions. The electrons are subject to radiative
processes in the flow, namely Bremsstrahlung, synchrotron, and
Comptonization of synchrotron photons. Moreover, we consider Coulomb
interaction between electrons and ions, which acts so as to heat the
electrons and cool the ions. Finally, we also consider the possibility
that the electrons are heated as a result of the turbulent-heating
processes. With all these radiative processes taken into account, we have
investigated the impact of their inclusion on the accretion flow and how
they influence the correlation between the temperatures of the different
components. The main results of the present work can be summarised as
follows:

\begin{itemize}
    \item [(1)] The inclusion of the Coulomb interaction and of the
      radiative cooling impacts the thermodynamical properties of both
      the ions and electrons. While the turbulent heating is effective
      only in the funnel and jet-sheath regions, the Coulomb interaction
      heats the electrons even in the high-density regions of the
      flow. Because these processes can change the temperature distribution 
      of the electrons significantly, and the latter is closely
      related to the electromagnetic emission from these accretion flows,
      our results underline the importance of a two-temperature approach
      in studying accretion flows around black holes and their
      imaging.\smallskip
    
    \item[(2)] The accretion rate is a key driver that influences the
      bulk properties of the flow and the thermodynamics of the electrons
      and ions. The density of the electrons and ions increases
      with the accretion rate, which, in turn, increases the
      radiation-cooling efficiency and the Coulomb-interaction rate. As a
      result, an increase of the accretion rate leads to electrons that
      are colder near the black hole but hotter far from it. \smallskip

    \item[(3)] We observe qualitatively distinct temperature properties
      for SANE and MAD accretion modes while maintaining the same
      accretion rates. In particular, since in the MAD mode the accreted
      magnetic flux is larger than in the SANE, the amount of synchrotron
      radiation produced is larger for a given accretion rate. In
      addition, because the turbulent heating dominates in the funnel and
      jet-sheath regions, we observe a hotter polar region and a cooler
      disc region. These differences may become more prominent when the
      accretion flow is imaged, and they could help distinguishing
      between MAD and SANE accretion flows via observations. We
        should note that the comparison between SANE and MAD models is
        here limited to a single accretion rate; future work will also
        explore additional MAD models with different accretion
        rates.\smallskip

    \item[(4)] Having to deal with two-temperature fluids clearly
      requires a different treatment of the temperatures of the
      electrons, the ions, and of the gas. The temperature ratios
      $T_i/T_e$ and $T_g/T_e$ show a broad correlation with the
      plasma-$\beta$ parameter\footnote{In principle, the temperature
      ratios could also be expressed in terms of the strength of the
      magnetisation $\sigma$. However, since the pressure and rest-mass
      density are related via an equation of state, the functional form
      of these alternative relations is expected to be rather similar.}
      and can therefore be expressed in terms of it. The standard
      $R-\beta$ relation \citep{Moscibrodzka-etal2016} fits well these
      correlations, but only for very small or very large values of the
      plasma-$\beta$ parameter, i.e., $\beta \lesssim 0.1$ and $\beta
      \gtrsim 100$. Because outside of these ranges, the $R-\beta$
      relation does not provide a good description of the temperature
      ratios well, we have modified the $R-\beta$ relations by
      introducing only two additional terms despite having to deal with
      two-temperature ratios. The new $R-\beta$ relations show a good fit
      of the correlations between the various temperatures in all the
      physical scenarios. \smallskip
    
    \item[(5)] The parameters involved in our new $R-\beta$ relations
      show a clear and strong correlation with the accretion rate. In
      particular, they are almost constant for very small accretion
      rates, but drastic variations in size when the accretion rate
      becomes higher than $\dot{M} \gtrsim
      10^{-7}\,\dot{M}_{\rm Edd}$. This result, which is in agreement
      with earlier studies \citep[][]{Dibi-etal2012, Ryan-etal2017,
        Yoon-etal2020}, suggests that a self-consistent magnetised
      two-temperature accretion flow is necessary to understand accretion
      flows with large accretion rates. \smallskip

\end{itemize}

In summary, our simulation models show that the two-temperature paradigm
is important for understanding the thermodynamical properties of
electrons and ions in the accretion flow. In particular, it is much more
critical if the accretion rate of flow is greater than
$\dot{M}\gtrsim10^{-7}\, \dot{M}_{\rm Edd}$. These conditions are those
expected to accompany the accretion onto supermassive black holes in
low-luminosity active galactic nuclei, where the accretion is expected to
be geometrically thick, optically thin, and radiatively inefficient
(RIAF) \citep[see][etc.]{Ichimaru1977, Rees-etal1982, Reynolds-etal1996,
  Narayan-etal1998, Remillard-McClintock2006}. Despite the fact that
  we only considered one black-hole mass for reference in this study,
  we account for radiative losses in code units, which are normalised by
  black-hole mass. We also express the accretion rate in terms of the
  Eddington accretion rate, and therefore the scaling of black-hole mass
  is taken into account naturally. Hence, our self-consistent
two-temperature approach can be used to model the near-horizon emission
from supermassive black holes such as Sgr A*, M\,87*. Such an
application, however, will necessarily require the extension of these
simulations to three spatial dimensions so as to guarantee the long-term
operation of the MRI and thus avoid the artificial decay of the magnetic
field that inevitably accompanies axisymmetric simulations. As a
  result, we anticipate that the results in the low-magnetised region
  ($\sigma<1$) of full 3D GRMHD simulations will be the same as those of
  our axisymmetric simulations. However, the results from our simulations
  might not be as realistic as those from 3D GRMHD simulations in the
  funnel region ($\sigma>1$) and may be slightly different
  quantitatively.

Generally, the electron temperature distribution or the distribution
  of the electron-to-ion temperature ratio $(T_e/T_i)$ also quite similar
  to the previous studies \citep[e.g.,][]{Ressler-etal2015,
    Ressler-etal2017,Sadowski-etal2017,Chael-etal2018,Chael-etal2019,
    Ryan-etal2018,Dexter-etal2020a,Mizuno-etal2021,Yao-etal2021}, apart
  from certain quantitative differences. Firstly, due to the turbulent
  heating prescription by \cite{Howes2010,Howes2011}, the electron
  temperature at the funnel region is very high and the ratio becomes
  $T_e/T_i>1$ in the MAD case. Similar results were reported by earlier
  studies on sub-grid electron heating \citep[e.g.,][]{Chael-etal2019,
    EHTCVIII}. Secondly, in the initial condition of our SANE models, we
  choose a smaller torus in comparison with those considered in earlier
  simulation work. This fact leaves our SANE models to have
  accumulated magnetic flux at the event horizon of the order of
  $\dot{\Phi}/\sqrt{\dot{M}} \sim1.5$. As a result, the electrons in the
  funnel region remain colder as compared with other studies
  \citep[e.g.,][]{Sadowski-etal2017,
    Dexter-etal2020a,Yao-etal2021}. Unlike most of the previous works, we
  do not restrict ourselves to studying the temperatures $(T_e, T_i)$
  distributions with different heating prescriptions, but we study the
  impact of the change in accretion rates by several orders of
  magnitudes.  And further, we go one step forward to compute the
  correlation of $T_e/T_i$ with the accretion rates. In the future, we
  plan to do 3D GRMHD simulations at different accretion rates and apply
  them to the recent observations of M\,87*, and Sgr A*.

In the governing Eq. \ref{eq:02}, we consider that the radiation
  flux term is proportional to the radiative cooling rate and the fluid
  velocity. This approximation is straightforward and computationally
  efficient to implement to study the impacts of radiation in the
  accretion flow \citep[e.g.,][etc.]{Fragile-Meier2009,Dibi-etal2012,
    Yoon-etal2020}. However, an M1 closure \citep{Sadowski-etal2017,
    Chael-etal2018, Chael-etal2019} or a Monte-Carlo scheme
  \citep{Ryan-etal2017,Ryan-etal2018,Dexter-etal2021} are more
  appropriate treatment for investigating the fluid dynamics coupling
  with radiation field, although it is much more computationally
  expensive. We plan to implement such schemes for our future studies. In
  this study, we chose the \cite{Howes2011} turbulent electron heating
  model, and for the sake of consistency, we kept it the same for all the
  simulation models with different physical scenarios considered
  here. Consideration of different electron heating models
  \citep[e.g.,][]{Rowan-etal2017,Werner-etal2018,
    Kawazura-etal2019,Zhdankin-etal2019} may calculate slightly different
  electron temperatures close to the black hole or the polar region
  \citep{Chael-etal2018,Dexter-etal2020, Mizuno-etal2021}. However, in
  the low-magnetised region $(\sigma<1)$, we do not anticipate any major
  differences in the obtained electron and ion
  temperatures. Additionally, we fixed the adiabatic index of the
  electrons to be $\tilde{\Gamma}_e=4/3$.  However, we observed thermally
  sub-relativistic electrons $(\Theta_e<1)$ in the cooling dominated
  region of the simulation domain. Electron temperatures calculated in
  these regions may be slightly over-estimated.  We should mention that
  using variable adiabatic index is more appropriate for studying
  magnetised accretion flow around black holes
  \citep[e.g.,][]{Sadowski-etal2017,Dihingia-etal2020}. We are currently
  working on such treatment, and we will report the results elsewhere in
  future.

\section*{Acknowledgements}

This research is supported by the DFG research grant ``Jet physics on
horizon scales and beyond" (Grant No. FR 4069/2-1), the ERC Advanced
Grant ``JETSET: Launching, propagation and emission of relativistic jets
from binary mergers and across mass scales'' (Grant No. 884631), Max
Planck partner group grant (MPG-01) at Indian Institute Technology of
Indore, the National Natural Science Foundation of China (Grant No. 12273022), and 
Department of Science and Technology's Swarnajayanti Fellowship 
(DST/SJF/PSA-03/2016-17) at IISc Bangalore. 
The simulations were performed on Pi2.0 and Siyuan Mark-I at
Shanghai Jiao Tong University, and Max Planck Gesellschaft (MPG)
super-computing resources. LR acknowledges the Walter Greiner
Gesellschaft zur F\"orderung der physikalischen Grundlagenforschung
e.V. through the Carl W. Fueck Laureatus Chair.
We appreciate the thoroughness and thoughtful comments provided 
by the reviewers that have improved the manuscript.

\section*{Data availability}

The data underlying this article will be shared on reasonable request to
the corresponding author.

\bibliographystyle{mnras}
\bibliography{references}


\appendix
\section{Correlation of best-fit parameters $\dot{M}$}
The values of the best-fit parameters for the temperature ratio $T_i/T_e~(R_{1-3,p}, \beta_{\rm br})$ for different simulation models are shown in Table \ref{tab-02}. To explore their correlations with the accretion rate, we find the best-fit of these parameters with the same. The explicit expressions of these parameters in terms of the accretion rate are given below, 
\begin{align}
    \begin{aligned}
    R_{1,p}(\dot{M}/\dot{M}_{\rm Edd}) =& R_{10} + R_{11}(\dot{M}/\dot{M}_{\rm Edd})^{0.54},\\
    R_{2,p}(\dot{M}/\dot{M}_{\rm Edd}) =& R_{20} + R_{21}(\dot{M}/\dot{M}_{\rm Edd})^{0.68},\\
    R_{3,p}(\dot{M}/\dot{M}_{\rm Edd}) =& R_{30} + R_{31}(\dot{M}/\dot{M}_{\rm Edd})^{0.41},\\
    \beta_{\rm br}(\dot{M}/\dot{M}_{\rm Edd}) =& \beta_{\rm br0} + \beta_{\rm br1}(\dot{M}/\dot{M}_{\rm Edd})^{0.16}.
    \end{aligned}
    \label{eq:rbcorr}
\end{align}
Here, the values of the different constants are given by $R_{10}=2.81,
R_{11}=-1.22\times10^2, R_{20}=1.60, R_{21}=4.14\times10^3, R_{30}=9.12,
R_{31}=-6.68\times10^2, \beta_{\rm br0}=62.4, \beta_{\rm br1}=
-2.50\times10^2$. In Fig. \ref{fig:para}, we show the plot of these
functions with dashed lines along with values of $(R_{1-3,p}, \beta_{\rm
  br})$ for different simulation models. The figure essentially shows
that the best-fit parameters correlate thoroughly with the mass-accretion
rate. Note that expressions~\eqref{eq:rbcorr} have been derived from
  SANE models only. Because the cooling and heating processes -- which
  are obviously the same for SANE and MAD accretion modes -- depend on
  the strength of the magnetic field, it is natural to expect that while
  the functional form will be the same, small changes in the values of
  the coefficients $R_{1,k},R_{2,k}, R_{3,k}$, and $\beta_{\rm br}$ will
  appear when evaluating expressions~\eqref{eq:rbcorr} for MAD accretion
  modes.


\bsp	
\label{lastpage}
\end{document}